\documentclass[lettersize,journal]{IEEEtran}
\usepackage{amssymb}
\usepackage{amsmath,amsfonts}
\usepackage{algorithmic}
\usepackage{algorithm}
\usepackage{array}
\usepackage[caption=false,font=normalsize,labelfont=sf,textfont=sf]{subfig}
\usepackage{textcomp}
\usepackage{stfloats}
\usepackage{url}
\usepackage{verbatim}
\usepackage{graphicx}
\usepackage{cite}
\begin{document}

\title{Quantum-inspired Multi-Parameter Adaptive Bayesian Estimation for Sensing and Imaging}

\author{Kwan Kit Lee, Christos N. Gagatsos, Saikat Guha,~\IEEEmembership{Senior Member, IEEE}, Amit Ashok,~\IEEEmembership{Member, IEEE}
\thanks{We acknowledge support for this work by the Defense Advanced Research Projects
Agency (DARPA) IAMBIC program under contract HR00112090128."}
}

\markboth{Journal of Selected Topics in Signal Processing, May~2022}%
{Shell \MakeLowercase{\textit{et al.}}: A Sample Article Using IEEEtran.cls for IEEE Journals}


\maketitle

\begin{abstract}
It is well known in Bayesian estimation theory that the conditional estimator ${\hat \theta} = E[\theta|l]$ attains the minimum mean squared error (MMSE) for estimating a scalar parameter of interest $\theta$ from  observations of $l$ through a noisy channel $P_{l|\theta}$, given a prior $P_\theta$ on $\theta$. In quantum, e.g., optical and atomic, imaging and sensing tasks the user has access to $\rho_\theta$, i.e. the quantum state that encodes $\theta$. The choice of a measurement operator, i.e. a positive-operator valued measure (POVM) $\Pi_l$, inducing the channel $P_{l|\theta} = {\rm Tr}(\rho_\theta \Pi_l)$, leads to a measurement outcome $l$, on which the aforesaid classical MMSE estimator is employed. Personick found the optimum POVM $\Pi_l$ that attains the MMSE over all possible physically allowable measurements and the resulting MMSE~\cite{Personick1971ApplicationOQ}. This result from 1971 is less-widely known than the quantum Fisher information (QFI), which lower bounds the variance of an unbiased estimator over all measurements without considering any prior probability. For multi-parameter estimation, in quantum Fisher estimation theory the inverse of the QFI matrix provides an operator lower bound on the covariance of an unbiased estimator, and this bound is understood in the positive semidefinite sense. However, there has been little work on quantifying the quantum limits and measurement designs, for multi-parameter quantum estimation in a {\em Bayesian} setting. In this work, we build upon Personick's result to construct a Bayesian adaptive (greedy) measurement scheme for multi-parameter estimation, when $N$ copies of $\rho_\theta$ are available. We illustrate our proposed measurement scheme with the application of localizing a cluster of point emitters in a highly sub-Rayleigh angular field-of-view, an important problem in fluorescence microscopy and astronomy. Our algorithm translates to a multi-spatial-mode transformation prior to a photon-detection array, with electro-optic feedback to adapt the mode sorter. We show that this receiver performs superior to quantum-noise-limited focal-plane direct imaging.
\end{abstract}

\begin{IEEEkeywords}
Quantum Information, Information Theory, Bayesian Inference, Super-Resolution. 
\end{IEEEkeywords}

\section{Introduction} \label{Sec:Intro}
In classical sensing and imaging paradigm, a measurement channel is modelled by a conditional probability $p(\boldsymbol l|\boldsymbol X(\boldsymbol\theta))$, where $\boldsymbol X(\boldsymbol\theta)$ and ${\boldsymbol l} = [l_{1}, l_{2},...,l_N]^T$ are the vector-valued measurement input (e.g., object/scene/signal) and outcome of the measurement channel respectively. The input $\boldsymbol X(\boldsymbol \theta)$ itself can be a deterministic function or a random variable parameterized by $M$ parameters $\boldsymbol\theta = [\theta_{1}, \theta_{2},...,\theta_{M}]^T$. Thus, the channel can be expressed by the conditional probability density $p(\boldsymbol l|\boldsymbol\theta)$. This measurement model can be also applied to quantum sensing, where the input $\boldsymbol X(\boldsymbol\theta)$ is replaced by a density operator $\rho({\boldsymbol \theta})$ describing the object being measured and the measurement channel is given by a positive-operator-valued measure (POVM) $\{\Pi_{\boldsymbol l}\}$ operating on $\rho({\boldsymbol \theta})$ resulting in outcome ${\boldsymbol l}$ with probability $p(\boldsymbol l|\boldsymbol\theta)={\rm Tr}\left(\rho({\boldsymbol \theta}) \Pi_{\boldsymbol l}\right)$~\cite{nielsen2001quantum}. Note that a classical measurement channel can always be expressed as: $\rho({\boldsymbol \theta})= \int p(\boldsymbol l|\boldsymbol\theta) d\boldsymbol l |\alpha_{\boldsymbol l}\rangle\langle\alpha_{\boldsymbol l}|$ with the projection operator POVM $\{\Pi_{\boldsymbol l}\} = \{|\alpha_{\boldsymbol l}\rangle\langle\alpha_{\boldsymbol l}|\}$, where $\{ |\alpha_{\boldsymbol l}\rangle \}$ is a set of orthonormal complete basis. Thus, in the following discussion we only consider the quantum formulation as the classical channel can be considered as a special case.  

\begin{figure}[h]
	\centering
	\includegraphics[width=0.45\textwidth]{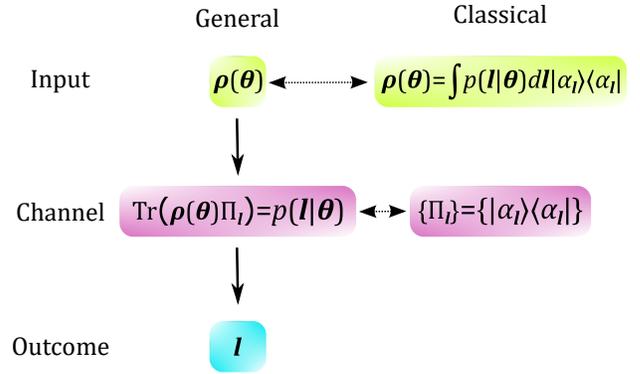}
	\caption{A schematic diagram shows the classical and quantum channel model.}
	\label{Channel}
\end{figure} 

\begin{figure}[h]
	\centering
	\includegraphics[width=0.45\textwidth]{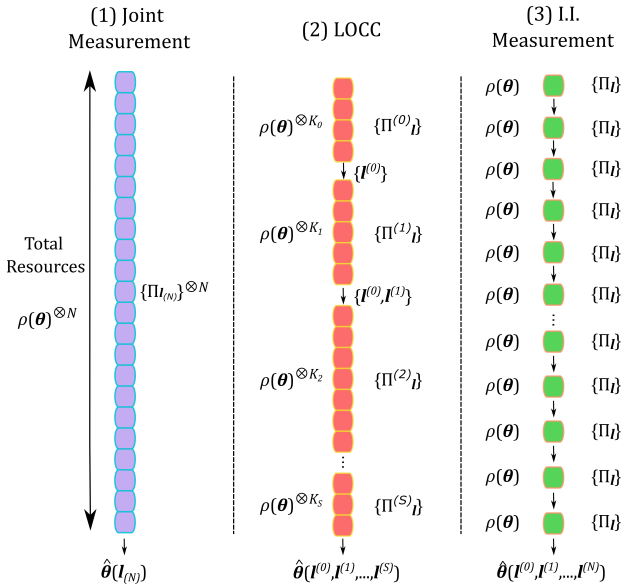}
	\caption{A schematic diagram shows the three different measurement approaches on $N$-copy of quantum states $\rho({\boldsymbol \theta})^{\otimes N}$.}
	\label{Measurements_general}
\end{figure}

In practice, if $N$ copies ($N \ge 2$) of quantum states $\rho({\boldsymbol \theta})^{\otimes N}$ are avaiable, the receiver can: (1) in the most general setting, choose a joint-measurement POVM $\{\Pi_{\boldsymbol l_{(N)}}\}$ acting collectively on $\rho({\boldsymbol \theta})^{\otimes N}$, producing the outcome ${\boldsymbol l_{(N)}}$; (2) employ the {\em local operations and classical communications} (LOCC) scheme, such that each batch of state $\rho({\boldsymbol \theta})^{\otimes K_{\tau}}$, where $K_{\tau}$ is the number of copies of state $\rho({\boldsymbol \theta})$ comprising the $\tau^{th}$ measurement batch, with $0\le\tau\le S$ and $N = \sum_{\tau=0}^{S} K_\tau$, is measured by the POVM $\{\Pi_{\boldsymbol l}^{(\tau)}\}$ chosen for example, based on the information available from the previous set of measurement outcomes \{${\boldsymbol l}^{(0)},{\boldsymbol l}^{(1)},\ldots, {\boldsymbol l}^{(\tau-1)}$\}; or (3) use independent identical measurements on each copy of the state, described by the POVM $\{\Pi_{\boldsymbol l}\}$. The schematic diagram illustrating these three measurement approaches is shown in Fig.~\ref{Measurements_general}. 

No matter the receiver strategy, after measuring all $N$ copies, the receiver generates an estimate of ${\boldsymbol \theta}$, i.e., $\hat {\boldsymbol \theta}(\boldsymbol l_{set})$ where $\boldsymbol l_{set} = {\boldsymbol l_{(N)}}$ for case (1) above, and $\boldsymbol l_{set} = [{\boldsymbol l}^{(0)},{\boldsymbol l}^{(2)},\ldots,{\boldsymbol l}^{(S)}]$ for cases (2) above and $\boldsymbol l_{set} = [{\boldsymbol l}^{(1)},{\boldsymbol l}^{(2)},\ldots,{\boldsymbol l}^{(N)}]$ for case (3) above. The receiver chooses the estimator to optimize a desired objective/loss function. A natural choice of the objective function associated with sensing and imaging estimation tasks is mean (expected) squared-error (MSE), ${\rm E}[||{\boldsymbol \theta} - {\hat {\boldsymbol \theta}}(\boldsymbol l_{set})||^2]$.

For any {\em given} measurement POVM $\{\Pi_{\boldsymbol l}\}$, assuming strategy (3) above, i.e., the same measurement acts on each copy of $\rho({\boldsymbol \theta})$, the problem reduces to the standard classical estimation theory problem of estimating ${\boldsymbol \theta}$ from $N$ i.i.d. samples of ${\boldsymbol l}$, each described by $p({\boldsymbol l} | {\boldsymbol \theta})$. The covariance $\text{Cov}(\hat{\boldsymbol \theta}(\boldsymbol l_{set}),{\boldsymbol \theta})$ for any unbiased estimator $\hat{\boldsymbol \theta}(\boldsymbol l_{set})$ of ${\boldsymbol \theta}$ is lower bounded by $\Sigma_{C}$. This means $\text{Cov}(\hat{\boldsymbol \theta}(\boldsymbol l_{set}),{\boldsymbol \theta})-\Sigma_{C}$ is a semi-positive definite matrix, denoted compactly as $\text{Cov}(\hat{\boldsymbol \theta}(\boldsymbol l_{set}),{\boldsymbol \theta}) \ge \Sigma_{C}$. The receiver's task is to pick the optimal estimator $\hat{\boldsymbol \theta}^{opt}(\boldsymbol l_{set})$ on the measurement outcomes $\boldsymbol l_{set}$, such that $\text{Cov}(\hat{\boldsymbol \theta}^{opt}(\boldsymbol l_{set}),{\boldsymbol \theta})$ saturates the bound $\Sigma_{C}$ when permissible.

Tools of quantum estimation theory allow us find a tight lower bound to $\text{Cov}(\hat{\boldsymbol \theta}(\boldsymbol l_{set}),{\boldsymbol \theta})$, which automatically optimizes over all physically-permissible choices of a POVM $\{\Pi_{\boldsymbol l}\}$ (again, assuming that the same measurement is used to detect each copy of $\rho({\boldsymbol \theta})$). The $\text{Cov}(\hat{\boldsymbol \theta}(\boldsymbol l_{set}),{\boldsymbol \theta})$  is lower bounded by $\Sigma_{Q}$ (a {\em quantum} bound), which itself is an infimum of all bounds $\Sigma_{C}$ associated with all possible choices of $\{\Pi_{\boldsymbol l}\}$. For certain cases (for example when ${\boldsymbol \theta}$ is a single scalar parameter), quantum estimation theory also provides the optimal receiver POVM $\{\Pi_{\boldsymbol l}^{(opt)}\}$. Once the optimal receiver is chosen, it uses the optimal estimator $\hat{\boldsymbol \theta}^{opt}(\boldsymbol l_{set})$ using standard classical estimation tools, such that covariance $\text{Cov}(\hat{\boldsymbol \theta}^{opt}(\boldsymbol l_{set}),{\boldsymbol \theta})$ saturates $\Sigma_Q$ when permissible. Therefore, in general we can state: $\text{Cov}(\hat{\boldsymbol \theta}(\boldsymbol l_{set}),{\boldsymbol \theta})\ge\Sigma_{C}\ge\Sigma_{Q}$, where $\Sigma_{C}$ corresponds to any choice of POVM. 

The aforementioned lower bounds on the covariance of multi-parameter estimators can be defined within the statistical inference frameworks of the frequentist approach, i.e., Fisherian (with no prior), or the Bayesian (with prior $p(\boldsymbol \theta)$) inference approach. We review below some known bounds for both inference approaches.

In the Fisherian (frequentist) approach, when no prior $p(\boldsymbol \theta)$ is available or defined, the Cramer-Rao lower bound (CRLB) $\Sigma_{C}$ on the covariance $\text{Cov}(\hat{\boldsymbol \theta}(\boldsymbol l),{\boldsymbol \theta})$ of an unbiased estimator is given by the inverse of the Fisher information (FI) matrix $I$~\cite{Kay97}:
\begin{align}
	I_{ij} = \int \bigg[ \frac{\partial}{\partial\theta_{i}}\ln p(\boldsymbol l | \boldsymbol\theta)\bigg] \bigg[ \frac{\partial}{\partial\theta_{j}}\ln p(\boldsymbol l |\boldsymbol\theta)\bigg]p(\boldsymbol l | \boldsymbol\theta) d\boldsymbol l, \label{CFI}
\end{align}
with $1 \le i,j \le M$, and the likelihood $p(\boldsymbol l | \boldsymbol\theta) = {\rm Tr}(\rho(\boldsymbol\theta) \Pi_{\boldsymbol l})$. The quantum version of this lower bound $\Sigma_{Q}$, which only depends on $\rho(\boldsymbol\theta)$ (since the measurement $\Pi_{\boldsymbol l}$ is automatically optimized over all POVMs) is given by the inverse of the quantum Fisher information (QFI) matrix $Q$~\cite{Liu_2019}, with elements:
\begin{align}
	Q_{ij} = {\rm Tr}\bigg[\rho(\boldsymbol\theta)\frac{L_iL_j+L_jL_i}{2}\bigg], \label{QFI}
\end{align}
where $L_i$ is the symmetric logarithmic derivative (SLD) operator. The SLD operator can be determined from the following implicit relationship:
\begin{align}
	2\frac{\partial}{\partial\theta_{i}}\rho(\boldsymbol\theta) = \rho(\boldsymbol\theta)L_i + L_i\rho(\boldsymbol\theta), \label{SLD}
\end{align}
with $1 \le i \le M$. Thus, we have $\text{Cov}(\hat{\boldsymbol \theta}(\boldsymbol l_{set}),{\boldsymbol \theta})\ge I^{-1} \ge Q^{-1}$ in the Fisher framework. For $N$-copy i.i.d. measurement of $\rho({\boldsymbol \theta})^{\otimes N}$, both the classical and quantum bounds scale by a factor of $1/N$. The classical one is asymptotically attained by the maximum likelihood estimator (MLE). The quantum CRLB ($Q^{-1}$) can not be saturated in general for $M > 1$.

The corresponding Bayesian lower bounds on the covariance $\text{Cov}(\hat{\boldsymbol \theta}(\boldsymbol l),{\boldsymbol \theta})$ of any estimator $\hat{\boldsymbol \theta}(\boldsymbol l)$ are found in~\cite{rubio2020bayesian}. Given a prior $p(\boldsymbol\theta)$ on the parameter vector $\boldsymbol\theta$, the Bayesian Cramer-Rao lower bound (BCRLB) $\Sigma_C$ is given by:
\begin{align}
	\Sigma_C = \int p(\boldsymbol\theta)\boldsymbol\theta\boldsymbol\theta^Td\boldsymbol\theta - J,\label{CBLB}
\end{align}
where the $M$-by-$M$ matrix $J$ is defined as:
\begin{align}
	J_{ij} = \int \frac{[\int p(\boldsymbol l,\boldsymbol\theta) \theta_{i}d\boldsymbol\theta][\int p(\boldsymbol l,\boldsymbol\theta) \theta_{j}d\boldsymbol\theta]}{p(\boldsymbol l)} d\boldsymbol l,\label{J_ele}
\end{align}
and $p(\boldsymbol l,\boldsymbol\theta) = p(\boldsymbol l | \boldsymbol\theta)p(\boldsymbol\theta)$ is the joint distribution of $\boldsymbol l$ and $\boldsymbol\theta$. The posterior mean of the parameters $\int \theta_{i} p(\boldsymbol\theta|\boldsymbol l) d\boldsymbol\theta$ saturate the bound in Eq.~(\ref{CBLB}). Further details about this estimator and the bound are described in Appendix~\ref{App_BCRB}.
For the quantum version of this lower bound, we first define the following operators, for $1 \le i \le M$ and $k = 0, 1, 2$~\cite{Personick1971ApplicationOQ}:
\begin{align}
\Gamma_{i,k} &= \int d\boldsymbol\theta  p(\boldsymbol\theta) \rho(\boldsymbol\theta) \theta_{i}^k, \label{Gamma} 
\end{align}
and operators $B_{i}$, $1 \le i \le M$, that satisfy:
\begin{align}
2\Gamma_{i,1} = \Gamma_{0}B_{i} + B_{i}\Gamma_{0}.\label{B_operator}
\end{align}	
For $k=0$, $\Gamma_{i,0} = \Gamma_{j,0}, \, \forall (i,j)$, thus we can drop the first index and denote it as $\Gamma_{0} = \int d\boldsymbol\theta  p(\boldsymbol\theta) \rho(\boldsymbol\theta)$, the average received state. The quantum BCRLB $\Sigma_Q$ can be written as:
\begin{align}
	\Sigma_Q = \int p(\boldsymbol\theta)\boldsymbol\theta\boldsymbol\theta^Td\boldsymbol\theta - G, \label{Sigma_Q}
\end{align}
where 
\begin{align}
	G_{ij} &= {\rm Tr}\bigg[\Gamma_{0}\frac{B_iB_j+B_jB_i}{2}\bigg]. \label{G_ele}	
\end{align}
Thus in a Bayesian inference framework, we have $\text{Cov}(\hat{\boldsymbol \theta}(\boldsymbol l),{\boldsymbol \theta})\ge \Sigma_C \ge \Sigma_Q$. Table~\ref{tab1} summarizes these Fisherian and Bayesian bounds.

\begin{table}[htbp]
	\caption{Summary of classical and quantum bounds for Fisherian and Bayesian approaches.}
	\begin{center}
		\begin{tabular}{|c|c|c|}
			\hline
			&\textbf{Fisher}&\textbf{Bayesian} \\
			\hline
			\textbf{Classical}& {Information Matrix: $I$} & {Information Matrix: $\Sigma_C$}\\
			\hline
			\textbf{Quantum} & {Information Matrix: $Q$} & {Information Matrix: $\Sigma_Q$}\\
			\hline
			\hline
			\textbf{Bounds} & $\text{Cov}(\hat{\boldsymbol \theta},{\boldsymbol \theta})\ge I^{-1}\ge Q^{-1}$ & $\text{Cov}(\hat{\boldsymbol \theta},{\boldsymbol \theta})\ge \Sigma_C \ge \Sigma_Q$\\
			\hline
		\end{tabular}
		\label{tab1}
	\end{center}
\end{table}

To achieve the quantum bound, an optimal measurement is required (i.e. an optimal choice POVM, that acts on each copy of $\rho({\boldsymbol \theta})$). For a single parameter problem ($M=1$), the projective measurement onto the eigenvectors of the SLD operator $L$ in Eq.~(\ref{SLD}) saturates the Fisher quantum bound, i.e., the $I$ for the SLD measurement equals $Q$. Likewise, the Bayesian quantum bound on the covariance is saturated (i.e. $\Sigma_C = \Sigma_Q$), for the case of a single parameter ($M=1$) by a projective measurement onto the eigenvectors of the operator $B$ in Eq.~(\ref{B_operator})~\cite{Personick1971ApplicationOQ}. 

For multi-parameter estimation, if the operators associated with parameter $\theta_i$: $L_i$ and $B_i$, $1 \le i \le M$ commute with one another, for the Fisher and Bayesian frameworks respectively, the corresponding covariance bound can be saturated by the above-said measurements, calculated by evaluating the eigenvectors of $L_i$ or $B_i$, respectively (which $i$ does not matter as they are simultaneously diagonal). However, if the operators do not commute, which is the case in general, a measurement that is jointly optimal for all parameters may not exist and/or likely to be challenging to derive. 

In the quantum case, the Holevo Cramer-Rao bound (HCRB)~\cite{HolevoBook} is the most
fundamental scalar lower bound on the weighted mean square error ${\rm Tr}[W \text{Cov}(\hat{\boldsymbol \theta}(\boldsymbol l),{\boldsymbol \theta})]$, for
a positive definite $W$. The HCRB represents the best precision attainable with a collective measurement (discussed as case (1) above) on an asymptotically large number of identical copies of $\rho({\boldsymbol \theta})$.

In this work, we propose a sequential adaptive (LOCC) measurement scheme for muti-parameter estimation within a full Bayesian inference framework by leveraging tools from the Bayesian quantum estimation theory. The details of our proposed measurement scheme are presented in Sec.~\ref{Bay_scheme}. In Sec.~\ref{Application}, we employ our measurement scheme to the problem of localizing an unknown number of point-emitters placed in a sub-Rayleigh (below diffraction-limit) field of view in an optical imaging context. This imaging application is motivated by the fact that traditional direct focal-plane imaging, which employs intensity measurements followed by electronic-domain processing, is known to be highly sub-optimal~\cite{PhysRevX.6.031033} in the sub-Rayleigh regime. We compare our quantum-inspired adaptive sequential measurement design with the direct imaging technique to quantify the significant optical resolution improvement obtained with our proposed scheme.

\section{Adaptive Sequential Measurement Scheme}\label{Bay_scheme}

Consider a system or a field in the state described by the density operator:
\begin{align}
	\rho(\boldsymbol\theta) = \sum_{i=1}^{P} b_{i}(\boldsymbol\theta)|\psi_{i}(\boldsymbol\theta)\rangle\langle \psi_{i}(\boldsymbol\theta)|, \label{do}
\end{align}
where $\boldsymbol\theta = [\theta_{1}, \theta_{2},...,\theta_{M}]^T$ are the parameters of interest,  $|\psi_{i}(\boldsymbol\theta)\rangle $ and $b_{i}(\boldsymbol\theta)$ are the parameter-dependent pure states and the corresponding weights respectively. As $\rho(\boldsymbol\theta)$ is unit trace, we have $\langle \psi_{i}(\boldsymbol\theta)|\psi_{i}(\boldsymbol\theta)\rangle = 1, \forall i$ and $\sum_{i=1}^{P} b_{i}(\boldsymbol\theta) = 1$. The states $|\psi_{i}(\boldsymbol\theta)\rangle $ are not necessarily orthogonal, i.e. $\langle \psi_{i}(\boldsymbol\theta)|\psi_{j}(\boldsymbol\theta)\rangle \neq 0$ for $i\neq j$ in general. $P$ itself, in general, is an unknown parameter (positive integer) such that: $P_{\rm min}\le P \le P_{\rm max}$. Here we assume that $P$ is upper bounded by $P_{\rm max}$, i.e., a prior on $P$. If the lower bound $P_{\rm min}$ is not known/available, we can set it to 1. When $P_{\rm min} \neq P_{\rm max}$, both $P$ and $\boldsymbol\theta$ need to be estimated. On the contrary, if $P_{\rm min} = P = P_{\rm max}$, i.e., $P$ is known \textit{a priori} exactly, then we only need to estimate the parameters $\boldsymbol\theta$.

\subsection{LOCC Measurement Scheme}\label{Single_Model}

\begin{figure}[h]
	\centering
	\includegraphics[width=0.45\textwidth]{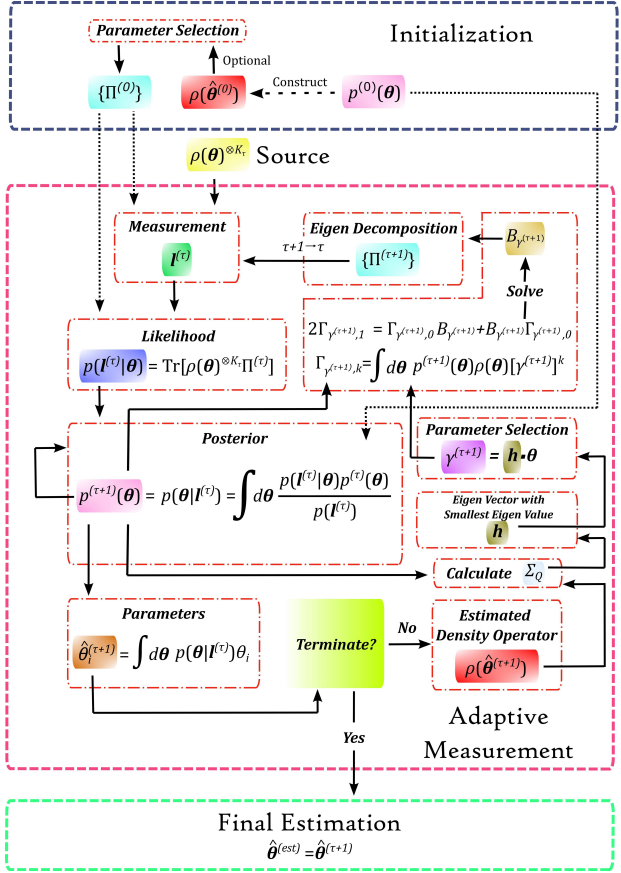}
	\caption{A schematic diagram illustrating various steps comprising our proposed sequential adaptive measurement scheme.}
	\label{Measurements_Bay_only}
\end{figure}

We design our measurement scheme within the LOCC framework to estimate multiple parameters $\boldsymbol\theta$ with $N$ independent copies of quantum state $\rho(\boldsymbol\theta)$ defined in Eq.~(\ref{do}). To illustrate our proposed scheme, we begin with the $P$ known exactly case. In the next section, we discuss an extension of this scheme where we relax this prior on $P$. The measurement scheme is illustrated in Fig.~\ref{Measurements_Bay_only}.

\subsubsection{Initialization}
The measurement is initialized by setting up $\{ \Pi^{{(0)}} \}$ and $p^{(0)}(\boldsymbol\theta)$, which are the POVM for measuring $\rho(\boldsymbol\theta)^{\otimes K_{0}}$ and the prior on the parameters $\boldsymbol\theta$ respectively. If by any means a set of pre-estimated parameters $\hat{\boldsymbol \theta}^{(0)}$ can be found, one may construct an estimated density operator $\rho(\hat{\boldsymbol \theta}^{(0)})$ and use the method described below to construct $\{ \Pi^{{(0)}}\}$. Otherwise, any POVM can be used in this stage.

\subsubsection{Measurement Cycle/Step}
Let us take $N = \sum_{\tau=0}^{S} K_\tau$, such that we adapt the measurement choice $S$ times, denoted by $\tau$ as the iteration index, $0\le\tau\le S$. In the $\tau^{th}$ measurement cycle, $K_{\tau}$ of copies of $\rho(\boldsymbol\theta)$, the density operator of which is $\rho(\boldsymbol\theta)^{\otimes K_{\tau}}$, are measured. The notation used here is the same as that in the previous section. In each measurement cycle/step, we employ the measurement strategy (3) introduced in Sec.~\ref{Sec:Intro}. Assume that in the $\tau^{th}$ measurement cycle, we have a POVM $\{ \Pi_{l^{(\tau)}} \}$ to measure \emph{each single copy} of $\rho(\boldsymbol\theta)$. For the $i^{th}$ copy of $\rho(\boldsymbol\theta)$, where $ 1\leq i\leq K_{\tau}$, the probability of obtaining the outcome $l_{i}^{(\tau)}$ is $p(l_i^{(\tau)}|\boldsymbol\theta) = {\rm Tr}[\rho(\boldsymbol\theta)\Pi_{l_i^{(\tau)}}]$, such that $\Pi_{l_i^{(\tau)}} \in \{ \Pi_{l^{(\tau)}} \}$. The probability of observing the measurement outcomes ${\boldsymbol l^{(\tau)}} = [l_{1}^{(\tau)}, l_{2}^{(\tau)},...,l_{K_{\tau}}^{(\tau)}]^T$ is $p(\boldsymbol l^{(\tau)}|\boldsymbol\theta) = {\rm Tr}[\rho(\boldsymbol\theta)^{\otimes K_{\tau}} \Pi^{(\tau)}] = \prod_{i=1}^{K_{\tau}} {\rm Tr}[\rho(\boldsymbol \theta) \Pi_{l_i^{(\tau)}}]$, where $\Pi^{(\tau)} \triangleq \Pi_{l_1^{(\tau)}} \otimes \ldots \otimes \Pi_{l_{K_{\tau}}^{(\tau)}}$. At the end of the sequential measurement scheme, a $N$-copy state $\rho({\boldsymbol \theta})^{\otimes N }$ has been measured. Note that  $K_{\tau}$ can be deterministic in some situations (e.g. the number of bits being transferred in a channel), but in many sensing/imaging problems, it is likely to be a random variable. For example, in the imaging problem discussed in the next section, a single photon is described by $\rho(\boldsymbol\theta)$ and the number of photons (copies of $\rho(\boldsymbol\theta)$) $K_{\tau}$ received in a fixed time period is a random variable governed by Poisson distribution. Nevertheless, our measurement protocol works for varying $K_{\tau}$ thus it fits naturally a wide range of sensing/imaging problems. 

The parameter estimate $\hat{\boldsymbol \theta}^{(\tau)}$, available after the $\tau^{th}$ sequential measurement is denoted by $\hat{\boldsymbol \theta}^{(\tau)} = [\hat{\theta}_{1}^{(\tau)}, \hat{\theta}_{2}^{(\tau)},...,\hat{\theta}_{M}^{(\tau)}]^T$. In a Bayesian inference setting, the  parameter estimate $\hat{\boldsymbol \theta}^{(\tau)}$ is given by posterior mean: $\hat{\boldsymbol \theta}^{(\tau)} = E_{p(\boldsymbol \theta|\boldsymbol l^{(\tau)})}[\boldsymbol \theta]$ if we wish to minimize the objective/loss function of MSE. For other loss functions (e.g., probability of detection/classification) other optimal estimators/detectors/classifiers can be chosen. The posterior is given by: $p(\boldsymbol \theta|\boldsymbol l^{(\tau)}) = p(\boldsymbol l^{(\tau)}|\boldsymbol \theta) \cdot p^{(\tau)}(\boldsymbol \theta) /p(\boldsymbol l^{(\tau)})$, where $p^{(\tau)}(\boldsymbol \theta)$ is the prior at the $\tau^{th}$ iteration. Note that the prior $p^{(\tau)}(\boldsymbol \theta)$ in turn equals the posterior $p(\boldsymbol \theta|\boldsymbol l^{(\tau-1)})$ at the previous $(\tau-1)^{th}$ iteration. The density operator at the $\tau^{th}$ iteration is represented as $\rho(\hat{\boldsymbol \theta}^{(\tau)})$. Now what remains to be determined is how we choose the POVM $\{\Pi_{l^{(\tau)}}\}$ at the $\tau^{th}$ iteration. We use the following strategy to pick/construct the POVM $\{\Pi_{l^{(\tau)}}\}$.

It is known that for a single parameter estimation problem, the eigen-projection measurement of $B_{1}$ in Eq.~(\ref{B_operator}) saturates the quantum bound $\Sigma_Q$~\cite{Personick1971ApplicationOQ}, which reduces to a lower bound of the variance of the scalar parameter. In this case the minimum mean square error (MMSE) is given by $\Sigma_Q={\rm Tr}[\Gamma_{1,2} - B_{1}\Gamma_{1,1}]$, where $\Gamma_{i,j}$ are defined in Eq.~(\ref{Gamma}). We refer to this measurement as the Personick projection in this work. For the multi-parameter problem, the counter-part of ${\rm Tr}[B_{1}\Gamma_{1,1}]$ is a matrix $G$ in Eq.~(\ref{G_ele}). If all $B_{i}$ operators commute, the quantum optimal measurement is given by the eigen-projections of any of the $B_{i}$ operators~\cite{rubio2020bayesian}. However, there is no such guarantee that the optimal measurement for all parameters exists or can be found in general. At the $\tau^{th}$ iteration of sequential measurement we define a single parameter $\gamma^{(\tau)}$, expressed as a linear combination of the $M$ parameters given by the the eigenvector of the matrix $\Sigma_Q$ with the smallest eigenvalue.  We claim that our approach is optimal (in MSE sense) for estimating a single parameter, which is linear combination of the multiple parameters of interest, in a given sensing/imaging problem. In Appendix~\ref{Para_sele}, we show the optimality of our single parameter estimation approach. Note that the matrix $\Sigma_Q$ is defined per Eq.~(9) for the density operator $\rho(\hat{\boldsymbol \theta}^{(\tau-1)})$. The scalar parameter $\gamma^{(\tau)}$ is used to construct the operator $B_{\gamma^{(\tau)}}$. The corresponding Personick projection constructed using $B_{\gamma^{(\tau)}}$ is chosen as the POVM $\{\Pi_{l^{(\tau)}}\}$ at the $\tau^{th}$ iteration.

The sequential measurements are terminated when all the $N$ available copies of $\rho(\boldsymbol\theta)$ have been exhausted.
\subsection{Extension: $P$ not known a priori}
If the scalar $P$ in Eq.~(\ref{do}) i.e. the number of parameters is unknown, we can employ and initialize multiple models of density operators $\rho(\boldsymbol\theta_P)$ with the corresponding prior $p(\boldsymbol\theta_P)$. Here $\boldsymbol\theta_P = [\theta_{1}, \theta_{2},...,\theta_{M_P}]^T$ for $ P_{min} \le P \le P_{max}$. In such a scenario, the number of parameters, denoted by $M_P$, for each model corresponding to a $P$ can be different in general. In $\tau^{th}$ iteration of the sequential measurement, one model is selected and used to construct the Personick measurement. The model can be selected randomly at $\tau = 0$, and the one that maximizes $p(\boldsymbol l^{(\tau-1)})$ can be used for the $\tau^{th}$ measurement iteration. We also propose an alternative model selection method in the next section. After model selection, the measurement scheme defined in the previous section can be applied unaltered. Note that at $\tau^{th}$ iteration, not only selected model but all the models are updated in a Bayesian inference setting, using the measurement outcome $\boldsymbol l^{(\tau)}$. When the sequential measurements eventually terminate, we can simply pick a model using the same model selection criteria described above and compute the final multi-parameter estimate as the posterior mean. However, other model selection criterias can also be applied as desired. 

\section{Application: Multi Point-emitter Estimation}\label{Application}
\subsection{Formulation}
We now illustrate our proposed adaptive sequential measurement scheme for estimating the location and relative brightness of incoherent point emitters comprising a cluster/constellation per the formulation in~\cite{PhysRevX.6.031033}. This type of estimation problem typically arises in many optical imaging applications such as astronomy and microscopy. The quantum state of photons incident on the image plane viewed through an optical lens is given by the density operator $\rho_{\text{full}}$:
\begin{align}
\rho_{\text{full}} &= (1-\epsilon)|\text{0}\rangle\langle\text{0}| + \epsilon\rho + O(\epsilon^2), \label{rho} 
\end{align} 
where $|\text{0}\rangle$ is the vacuum state, $\rho$ is the single photon state density operator, which has the form of Eq.~(\ref{do}), and $\epsilon$ is the total number of photons arriving on the image plane within the coherence time of the source. Assuming that $\epsilon\ll1$ (valid for weak thermal source), the photon states with order $O(\epsilon^2)$ are negligible. As the vacuum state $|\text{0}\rangle$ provides no information, we can focus on $\rho$. Thus, the components of Eq.~(\ref{do}) have the following meaning in the present problem context: $P$ is the number of point emitters, $\{b_i\}_{i=1}^{P}$ are the relative brightness of each point emitter or source (sum normalized to 1) and the states $|\psi_i\rangle$ are given by:
\begin{align}
|\psi_i\rangle &= \int_{-\infty}^{\infty}\int_{-\infty}^{\infty}\psi(x-x_i,y-y_i)|x,y\rangle dxdy , \label{psi_irangle} 
\end{align}
such that $(x_i,y_i)$ are the coordinates of the $i^{th}$ point source on the image plane. Here the point spread function (PSF) $\psi(x,y)$ of the imaging system is modeled by a 2D Gaussian function:
\begin{align}
\psi(x,y) &= \frac{1}{\sqrt{2\pi\sigma_x\sigma_y}}\exp\bigg(-\frac{x^2}{4\sigma_x^2}-\frac{y^2}{4\sigma_y^2}\bigg), \label{PSF} 
\end{align} 
where $\sigma_x$ and $\sigma_y$ are the standard deviation (a measure of width) of the PSF in $x$ and $y$ direction respectively. For a given PSF, $\sigma_x$ and $\sigma_y$ are known parameters and set to $\sigma_x=\sigma_y$ in our study. We define the full width at half maximum (proportional to $\sigma_x$) of the PSF as Rayleigh length (rl) in our analysis.

The parameters of interest in this problem are thus the position and relative brightness of the $P$ point emitters, i.e. $\boldsymbol\theta = [x_1,...,x_P,y_1,...,y_P,b_1,...,b_P]^T = [\boldsymbol x, \boldsymbol y, \boldsymbol b]^T$.

For the positions $[\boldsymbol x, \boldsymbol y]^T$, we use an independent Gaussian distribution $\mathcal{N}$ prior:
\begin{align}
p(\boldsymbol x,\boldsymbol y) &= \prod_{i}^{P} \mathcal{N}(x_i;\bar{x}_i,\bar{\sigma}_{x_i}) \mathcal{N}(y_i;\bar{y}_i,\bar{\sigma}_{y_i}),
\end{align}
where for $1\le i \le P$, $\bar{x}_i, \bar{y}_i, \bar{\sigma}_{x_i}, \bar{\sigma}_{y_i}$ are the mean and standard deviation of the position parameters $x_i$ and $y_i$ respectively. 

For the brightness $\boldsymbol b^T$ parameters a Dirichlet distribution~\cite{degroot1969optimal} is used as a prior: $p(\boldsymbol b) = \text{Dir}(\boldsymbol b;\boldsymbol a)$, where $\boldsymbol a = [a_1,...,a_P]^T$ are the hyper-parameters of the Dirichlet distribution. Thus, the overall prior is expressed as: $p(\boldsymbol x,\boldsymbol y, \boldsymbol b) = p(\boldsymbol x,\boldsymbol y)p(\boldsymbol b) $.

We have defined all relevant detail (i.e., photon state density operator, prior distribution) for the proposed adaptive sequential measurement scheme described in the previous section. Note that as $p(\boldsymbol x,\boldsymbol y, \boldsymbol b)$ is not a conjugate prior for the Poisson likelihood, we update the hyper-parameters of the prior distribution at  $\tau^{th}$ iteration to derive the posterior, which assumes the role of the prior in the next $(\tau+1)^{th}$ iteration. The prior hyper-parameters are: $\boldsymbol h=[\bar{x}_1,.., \bar{x}_P, \bar{y}_1,.., \bar{y}_P, \bar{\sigma}_{x_1},.., \bar{\sigma}_{x_P}, \bar{\sigma}_{y_1},.., \bar{\sigma}_{y_P}, a_1,..a_P, \delta]^T = [\bar{\boldsymbol x}, \bar{\boldsymbol y}, \bar{\boldsymbol \sigma}_{x}, \bar{\boldsymbol\sigma}_{y},  \boldsymbol a, \delta]^T$. Here, $\delta$ is another hyper-parameter associated with the brightness prior distribution which is explained later.

To update the hyper-parameters of the position prior at the $(\tau+1)^{th}$ iteration, we use the first- and the second-moments of the posterior distribution at the $\tau^{th}$ iteration:

\begin{align}
\bar{\alpha}_i^{(\tau+1)} &= \int \alpha_i p(\boldsymbol \theta|\boldsymbol l^{(\tau)}; \boldsymbol h^{(\tau)}) d\boldsymbol \theta, \label{mom_1} \\
\bar{\sigma}_{\alpha_i^{(\tau+1)}}^2 &= \int [\alpha_i - \alpha_i^{(\tau+1)}]^2 p(\boldsymbol \theta|\boldsymbol l^{(\tau)}; \boldsymbol h^{(\tau)}) d\boldsymbol \theta, \label{mom_2}
\end{align} 
where $\alpha$ represents $x$ or $y$ co-ordinate.

For the hyper-parameters $\boldsymbol a^T$ of the brightness prior, an expectation maximization (EM) approach is used. We first find the mean of the brightness vector as:

\begin{align}
\hat{b}_i^{(\tau+1)} = \int b_i p(\boldsymbol \theta|\boldsymbol l^{(\tau)}; \boldsymbol h^{(\tau)}) d\boldsymbol \theta.\label{b_max}
\end{align}
Then, $\boldsymbol a^T$ is updated such that $\hat{\boldsymbol b}^{(\tau+1)}$ becomes the mode of the distribution:

\begin{align}
\boldsymbol a^{(\tau+1)} &= \hat{\boldsymbol b}^{(\tau+1)}[ a_0^{(\tau)} + \delta^{(\tau)} - P ] + 1\nonumber \\
                         &= \hat{\boldsymbol b}^{(\tau+1)}[ a_0^{(\tau+1)} - P ] + 1, \label{a_update}
\end{align} 
where $a_0^{(\tau)} = \sum_{i}^{P}a_i^{(\tau)}$ and $a_0^{(\tau+1)} = a_0^{(\tau)} + \delta^{(\tau)}$. Qualitatively the larger the $a_0^{(\tau)}$, the smaller the total variance of the Dirichlet distribution. Adding  $\delta^{(\tau)} \ge 0$ leads to $a_0^{(\tau+1)} \ge a_0^{(\tau)}$, such that the variance reduces monotonically with each iteration $\tau$. Note that the introduction of $\delta^{(\tau)}$ does not change the position of the mode in the distribution. We set $\delta^{(\tau)}$ to a constant for all $\tau$.

When $P$ (i.e. number of point emitters) is unknown \textit{a priori}, we select the model in each measurement cycle as follow. Let $p_{P}(\boldsymbol l^{(\tau)})$ denote the likelihood of the model consisting of $P$ point emitters in the $\tau^{th}$ cycle. We calculate the following weighted log likelihood $Z_P^{(\tau)}$:
\begin{align}
Z_P^{(\tau)} &= \sum_{t=1}^{\tau} \exp\bigg[-\kappa\bigg( 1-\frac{t}{\tau} \bigg) \bigg] \ln p_{P}(\boldsymbol l^{(t)}),
\end{align} 
and pick the model with largest $Z_P^{(\tau)}$ as the estimate in the $(\tau+1)^{th}$ measurement cycle.
\subsection{Simulation Results}
We demonstrate the performance of the proposed adaptive sequential measurement scheme for 100 distinct realizations of 3-point emitter constellations. 

The position of the $1^{st}$ point emitter is uniformly distributed inside a circle with radius of 0.375 rl (Rayleigh length). The position of the $i^{th}$ emitter, $i > 1$, is $[x_{i}, y_{i}] = [x_{i-1}, y_{i-1}] + [(d+\delta d){\rm cos}\phi, (d+\delta d){\rm sin}\phi]$, where $d$ is a constant, $\delta d$ and $\phi$ are uniformly distributed random variables over the intervals $[-\delta d_0/2,\delta d_0/2]$ and $[0,2\pi)$ respectively, for some constant $\delta d_0$, such that $0\leq\delta d_0 < 2d$. The position of the $i^{th}$ emitter $[x_{i}, y_{i}]$ is re-selected if it falls outside the 0.375 rl circle (field of view) or the separation of any pair of sources is smaller than $d-\delta d_0/2$. By doing so, for each emitter, the closest neighbour is located around $d$ and minimum separation of any pair of point emitters is guaranteed to be no less than $d-\delta d_0/2$. In the simulation below we set $d=0.1$ (rl) and $\delta d_0 = 0.1 d$. The relative brightness of point emitters set to be equal/uniform. The average total photon budget $N$ is set to $5\times10^5$ and each adaptive sequential step utilizes around $10^4$ photons (i.e. the mean of $K_{\tau}$ is $10^4$ for $\tau \leq$ 1). The adaptive sequential scheme is initialized by employing 1000 photons for a direct imaging measurement (i.e. the mean of $K_{0}$ is 1000) followed by using an expectation maximization (EM) algorithm to estimate the initial model parameters. The remaining photons are detected by using Personick projection measurement in each adaptive sequential step. 

For the traditional direct imaging (serves as a baseline), which uses direct focal plane intensity measurements of all available N photon copies, the Richardson-Lucy deconvolution algorithm~\cite{Richardson:72} is first used to deconvolve the blurred image followed by the k-mean clustering algorithm~\cite{kodinariya2013review} to find the position and relative brightness of identified point emitters. 

\subsection{Estimation with $P$ known exactly}

\begin{figure}[h]
	\centering
	\includegraphics[width=0.45\textwidth]{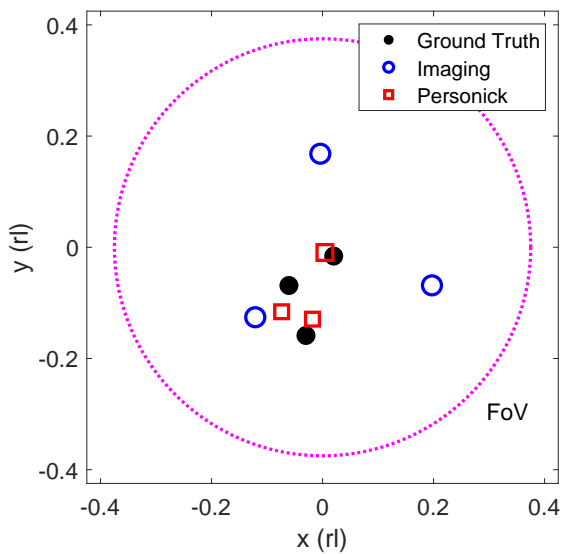}
	\caption{An illustrative example of a cluster of point emitter estimated with direct imaging (baseline) and Personick projection measurements (proposed adaptive measurement), when $P$ is \emph{known} exactly. The black dots, blue circles and red squares correspond to the ground truth, estimates obtained from direct imaging and Personick projection measurements respectively. The marker size is proportional to the point emitter brightness.}
	\label{Est_example}
\end{figure} 

For each of the 100 constellations, we employ 10 Monte Carlo simulation (i.e. different noise realizations). Fig.~\ref{Est_example} shows an illustrative realization of the point emitter cluster and estimated location and brightness using the two measurement schemes. 

\begin{figure}[h]
	\centering
	\includegraphics[width=0.45\textwidth]{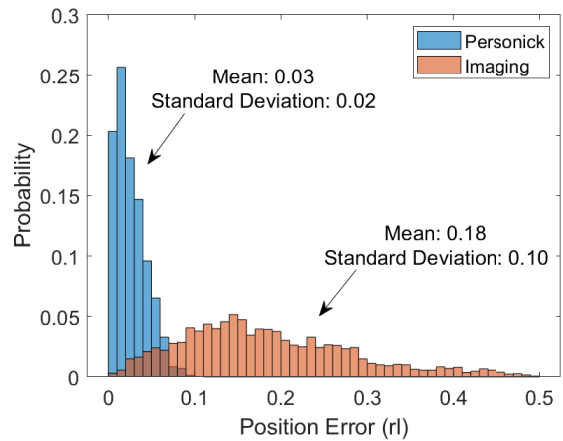}
	\caption{Distribution of the point emitter position errors obtained with the two measurement schemes, when $P$ is known exactly.}
	\label{N_knownos}
\end{figure}

To obtain the average performance of the proposed adaptive measurement scheme, for each point emitter realization, we first pair the ground truth point emitter location with the estimated locations, such that the sum of the position errors defined as: $\sum_{i=1}^{P}\sqrt{ (x_i-\hat{x_i})^2 + (y_i-\hat{y_i})^2}$, over all point-source matched pairs is minimized. The average (over all emitters) position error distribution of the point emitters is shown in Fig.~\ref{N_knownos}. We observe that the proposed adaptive scheme outperforms the direct imaging. More specifically, the mean position error obtained by the adaptive scheme is six-fold lower than that of the direct imaging. Also, the position error distribution of the Personick measurement is more concentrated and position errors for all estimates is less than $d=0.1$(rl).

\subsection{Estimation with unknown $P$}
\begin{figure}[h]
	\centering
	\includegraphics[width=0.45\textwidth]{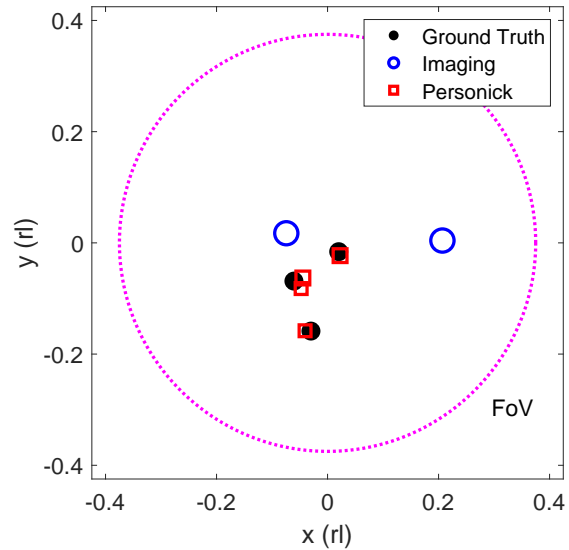}
	\caption{An illustrative example of a cluster of point emitter estimated with direct imaging and Personick projection measurements, when $P$ is \emph{unknown}. The marker definition is same as in Fig.~4.}
	\label{Est_example_N_unknown}
\end{figure}
\begin{figure}[h]
	\centering
	\includegraphics[width=0.45\textwidth]{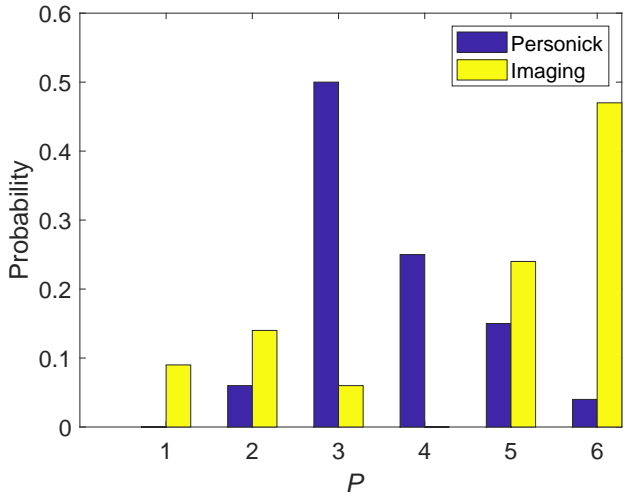}
	\caption{Distribution of the number of point emitters estimated by the two measurement schemes.}
	\label{N_est}
\end{figure}

\begin{figure}[h]
	\centering
	\includegraphics[width=0.45\textwidth]{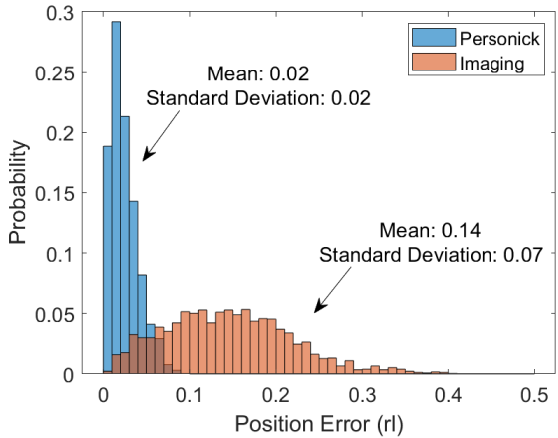}
	\caption{Distribution of the point emitter position errors obtained with the two measurement schemes, when $P$ is not known \textit{a priori}.}
	\label{N_unknownos}
\end{figure}

 
When $P_{max}=6$ is used as a prior, the estimation algorithm has to also estimate $P$. One of such illustrative example is shown in Fig.~\ref{Est_example_N_unknown}. It can be observed that even if the Personick measurement predicts the wrong number of sources (i.e. four instead of three), the reconstructed point emitter distribution closer to the ground truth compared to the reconstruction obtained with the direct imaging measurement, which underestimated the number of point emitters as two in this particular instance. To analyze the performance quantitatively, using the same set of constellations and same number of simulations, the distribution of number of point emitters estimated by the two measurement schemes in shown in Fig.~\ref{N_est}. We observe that the adaptive Personick projective scheme estimates the correct number of point emitters with a 50\% success rate relative to only 10\% for direct imaging. The $P$ estimated by our proposed Personick projective measurement scheme is more concentrated around $P=3$ while that of direct imaging is more spread out across the range of possible $P$. Fig.~\ref{N_unknownos} shows the corresponding position error distribution, computed only for cases where $P\geq 3$ in which none of the estimated point emitters sources are merged. We observe that when the $P$ is estimated correctly the proposed adaptive scheme maintains the significant performance advantage over direct imaging in terms of lower point emitter localization error.

\section{Conclusions}
Based on quantum estimation theory, we propose an adaptive sequential Bayesian multi-parameter estimation scheme with applications in sensing and imaging. Using an illustrative example task of optical super-resolution of point emitters embedded in a constellation, relevant in many optical imaging applications such as astronomy and microscopy, we demonstrate its superior performance relative to the direct diffraction-limited imaging approach. Our simulation study results show a nearly six-fold lower point emitter localization error achieved by our proposed measurement/estimation scheme relative to direct imaging scheme in the sub-Rayleigh regime. It is also worth noting that our POVM choice i.e. measurement design used in each sequential measurement step is provably optimal (in the MSE sense) when estimating a single scalar parameter, which is a linear combination of the multiple parameters of interest in a given estimation task. We believe our proposed multi-parameter estimation scheme is an important step towards achieving quantum optimal performance for practical sensing and imaging tasks, especially for sources/objects/signals that are fully described by semi-classical models that span a wide-range of real-world applications, such as exo-planet search, fluorescence microscopy and space situational awareness. We are currently exploring extensions of proposed multi-parameter estimation scheme along various directions; including optimal or near-optimal measurement design for more than one parameters in each adaptive sequential step as as well as considering non-greedy adaptive sequential scheme(s) inspired by our prior work~\cite{Huang16}. 

\appendices
\section{Posterior Mean Saturates Quantum Bound}\label{App_BCRB}
For any parameters $\boldsymbol\theta = [\theta_{1}, \theta_{2},...,\theta_{M}]^T$ and their estimators $\hat{\boldsymbol\theta}(\boldsymbol l) = [\hat{\theta}_{1}(\boldsymbol l), \hat{\theta}_{2}(\boldsymbol l),...,\hat{\theta}_{M}(\boldsymbol l)]^T$, where ${\boldsymbol l} = [l_{1}, l_{2},...,l_N]^T$ are the measurement through a channel represented by POVM $\{ \Pi \}$, the covariance matrix elements are: $[\text{Cov}(\hat{\boldsymbol \theta}(\boldsymbol l),{\boldsymbol \theta})]_{ij} = {\rm E}[(\theta_i - \hat{\theta}_{i}(\boldsymbol l))(\theta_j - \hat{\theta}_{j}(\boldsymbol l))]$. In Bayesian setting, the expectation is taken over the joint distribution $p(\boldsymbol\theta, \boldsymbol l)$ of both $\boldsymbol\theta$ and $\boldsymbol l$. Upon expending, the covariance matrix elements can be also written as:
\begin{align}
    [\text{Cov}(\hat{\boldsymbol \theta}(\boldsymbol l),{\boldsymbol \theta})]_{ij} &= {\rm E}[\theta_i \theta_j] - {\rm E}[\theta_i\hat{\theta}_{j}(\boldsymbol l)] - {\rm E}[\theta_j\hat{\theta}_{i}(\boldsymbol l)] \nonumber \\
    &+ {\rm E}[\hat{\theta}_{i}(\boldsymbol l)\hat{\theta}_{j}(\boldsymbol l)]. \label{Cov_exp}
\end{align}
The first term ${\rm E}[\theta_i \theta_j]$ in Eq.~(\ref{Cov_exp}) matches the first term in Eq.~(\ref{CBLB}). Now, substitute $\hat{\theta}_{i}(\boldsymbol l) = \int \theta_i p(\boldsymbol \theta|\boldsymbol l)d\boldsymbol \theta$, the second term of Eq.~(\ref{Cov_exp}) becomes:
\begin{align}
    {\rm E}[\theta_i\hat{\theta}_{j}(\boldsymbol l))] &= \int\int \theta_i \bigg[\int \theta'_j p(\boldsymbol \theta'|\boldsymbol l) d\boldsymbol \theta' \bigg] p(\boldsymbol \theta, \boldsymbol l) d\boldsymbol \theta d\boldsymbol l \nonumber \\
    &= \int \bigg[\int \theta_i p(\boldsymbol \theta|\boldsymbol l) d\boldsymbol \theta \bigg] \bigg[\int \theta'_j p(\boldsymbol \theta'|\boldsymbol l) d\boldsymbol \theta' \bigg] p(\boldsymbol l) d\boldsymbol l \nonumber \\ 
    &= J_{ij}.\label{Cov_exp_2nd}
\end{align}
Similarly, the third and the fourth terms in Eq.~(\ref{Cov_exp}) equal $J_{ij}$, and thus $\text{Cov}(\hat{\boldsymbol \theta}(\boldsymbol l),{\boldsymbol \theta})=\Sigma_C$ exactly. If $M=1$, i.e. $\boldsymbol \theta$ is a single parameter, $\Sigma_C$ gives the minimum mean square error with the posterior mean as the estimator.

\section{Parameter Selection}\label{Para_sele}
For a single parameter $\gamma = \vec{h}\cdot\vec{\theta}$, where $\vec{h}$ is some unit vector, the variance of estimating $\gamma$ is:
\begin{align}
\text{Var}(\gamma) &= \int\int d\vec{\theta}d\vec{l} [\vec{h}\cdot\hat{\theta}(\vec{l})-\vec{h}\cdot\vec{\theta}]^2 \nonumber \\
 &= \vec{h}^T\text{Cov}[\hat{\theta}(\vec{l}),\vec{\theta}]\vec{h}  \nonumber\\
 &\geq \vec{h}^T\Sigma_{Q}\vec{h}. \label{Personick_bound_gamma}
\end{align}
Applying Eq.~(\ref{Gamma}) for $k=1$ to $\gamma$, we have:
\begin{align}
\Gamma_{1,\gamma} &= \int d\vec{\theta} p(\vec{\theta}) \rho(\vec{\theta}) (\vec{h}\cdot\vec{\theta}) \nonumber\\
&= \sum_i h_i \Gamma_{1,i}. \label{Gamma_gamma}
\end{align}
Thus, if we set $B_{\gamma} = \sum_i h_i B_{i}$, Eq.~(\ref{B_operator}) can be satisfied. Under the constrain $|h|^2 = 1$, we search for the $\vec{h}$ that minimizes the MMSE$={\rm Tr} [\Gamma_{2,\gamma}-B_\gamma\Gamma_{1,\gamma}]$ using Lagrange multiplier:
\begin{align}
L &= \int d\vec{\theta} p(\vec{\theta}) (\vec{h}\cdot\vec{\theta})^2 - \text{Tr}\bigg[ \bigg(\sum_j h_j B_{j}\bigg) \bigg(\sum_k h_k \Gamma_{1,k}\bigg)\bigg] \nonumber\\
&- \lambda\bigg(\sum_i h_i^2 - 1\bigg).
\end{align}
Taking the derivative with respect to $\vec{h}$ and $\lambda$, we have: 
\begin{align}
\frac{\partial L}{\partial h_i} &=  2\int d\vec{\theta} p(\vec{\theta})(\vec{h}\cdot\vec{\theta})\theta_i - 2\lambda h_i \nonumber\\
&-\text{Tr}\bigg[ B_{i} \bigg(\sum_k h_k \Gamma_{1,k}\bigg) + \bigg(\sum_j h_j B_{j}\bigg)  \Gamma_{1,i}\bigg] = 0, \label{d_L_d_hi}\\
\frac{\partial L}{\partial \lambda} &= \sum_i h_i^2 - 1 = 0 \label{d_L_d_lambda}.
\end{align}
Using Eq.~(\ref{d_L_d_hi}), we have:
\begin{align}
h_i &= \frac{1}{\lambda} \sum_j h_j \bigg[ \int d\vec{\theta} p(\vec{\theta}) \theta_i\theta_j - \frac{1}{2}\text{Tr}\bigg( B_{i}\Gamma_{1,j} + B_{j}\Gamma_{1,i}\bigg) \bigg]\nonumber\\
    &= \frac{1}{\lambda} \sum_j h_j \bigg[ \int d\vec{\theta} p(\vec{\theta}) \theta_i\theta_j - \text{Tr}\bigg( \Gamma_{0} \frac{B_{i}B_{j}+B_{j}B_{i}}{2} \bigg) \bigg]\nonumber\\
    &= \frac{1}{\lambda} \sum_j h_j \bigg[ \int d\vec{\theta} p(\vec{\theta}) \theta_i\theta_j - G_{ij} \bigg].\label{L_hi}
\end{align}
Substituting Eq.~(\ref{L_hi}) into Eq.~(\ref{d_L_d_lambda}), we have:
\begin{align}
\lambda &= \sqrt{\sum_{i}\bigg\{\bigg[ \sum_j h_j \bigg[ \int d\vec{\theta} p(\vec{\theta}) \theta_i\theta_j - G_{ij} \bigg]\bigg\}^2}. \label{L_lambda}
\end{align}
It can be easily seen that Eq.~(\ref{L_hi}) is precisely the equations to determine the eigenvector of $\Sigma_{Q}$, with the corresponding eigenvalue $\lambda$. Thus, if we pick the eigenvector $\vec{h}_m$ with the smallest eigenvalue $\lambda_m$, the MMSE would be: $\vec{h}_m^T\Sigma_{Q}\vec{h}_m = \lambda_m$, which can be saturated by the eigen-projection measurement of $B_\gamma$. Since the MMSE is lower bounded by 0, $\lambda_m$ is the global minimum. In other words, $\lambda_m$ is the minimum MMSE one can get for a single parameter which is the linear combination of the original parameters $\vec{\theta}$. 

\bibliographystyle{IEEEtran}
\bibliography{IEEEabrv,sample}

\begin{thebibliography}{10}
\providecommand{\url}[1]{#1}
\csname url@samestyle\endcsname
\providecommand{\newblock}{\relax}
\providecommand{\bibinfo}[2]{#2}
\providecommand{\BIBentrySTDinterwordspacing}{\spaceskip=0pt\relax}
\providecommand{\BIBentryALTinterwordstretchfactor}{4}
\providecommand{\BIBentryALTinterwordspacing}{\spaceskip=\fontdimen2\font plus
\BIBentryALTinterwordstretchfactor\fontdimen3\font minus
  \fontdimen4\font\relax}
\providecommand{\BIBforeignlanguage}[2]{{%
\expandafter\ifx\csname l@#1\endcsname\relax
\typeout{** WARNING: IEEEtran.bst: No hyphenation pattern has been}%
\typeout{** loaded for the language `#1'. Using the pattern for}%
\typeout{** the default language instead.}%
\else
\language=\csname l@#1\endcsname
\fi
#2}}
\providecommand{\BIBdecl}{\relax}
\BIBdecl

\bibitem{Personick1971ApplicationOQ}
S.~D. Personick, ``Application of quantum estimation theory to analog
  communication over quantum channels,'' \emph{IEEE Trans. Inf. Theory},
  vol.~17, pp. 240--246, 1971.

\bibitem{nielsen2001quantum}
M.~A. Nielsen and I.~L. Chuang, ``Quantum computation and quantum
  information,'' \emph{Phys. Today}, vol.~54, no.~2, p.~60, 2001.

\bibitem{Kay97}
S.~M. Kay, \emph{Fundamentals of Statistical Signal Processing: Estimation
  Theory}.\hskip 1em plus 0.5em minus 0.4em\relax Prentice Hall, 1997.

\bibitem{Liu_2019}
\BIBentryALTinterwordspacing
J.~Liu, H.~Yuan, X.-M. Lu, and X.~Wang, ``Quantum fisher information matrix and
  multiparameter estimation,'' \emph{Journal of Physics A: Mathematical and
  Theoretical}, vol.~53, no.~2, p. 023001, dec 2019. [Online]. Available:
  \url{https://doi.org/10.1088/1751-8121/ab5d4d}
\BIBentrySTDinterwordspacing

\bibitem{rubio2020bayesian}
J.~Rubio and J.~Dunningham, ``Bayesian multiparameter quantum metrology with
  limited data,'' \emph{Physical Review A}, vol. 101, no.~3, p. 032114, 2020.

\bibitem{HolevoBook}
A.~Holevo, \emph{Probabilistic and Statistical Aspects of Quantum
  Theory}.\hskip 1em plus 0.5em minus 0.4em\relax Edizioni della Normale, 2011.

\bibitem{PhysRevX.6.031033}
\BIBentryALTinterwordspacing
M.~Tsang, R.~Nair, and X.-M. Lu, ``Quantum theory of superresolution for two
  incoherent optical point sources,'' \emph{Phys. Rev. X}, vol.~6, p. 031033,
  Aug 2016. [Online]. Available:
  \url{https://link.aps.org/doi/10.1103/PhysRevX.6.031033}
\BIBentrySTDinterwordspacing

\bibitem{degroot1969optimal}
\BIBentryALTinterwordspacing
M.~DeGroot and M.~DEGROOT, \emph{Optimal Statistical Decisions}, ser.
  McGraw-Hill series in probability and statistics.\hskip 1em plus 0.5em minus
  0.4em\relax McGraw-Hill, 1969. [Online]. Available:
  \url{https://books.google.com/books?id=39UznQEACAAJ}
\BIBentrySTDinterwordspacing

\bibitem{Richardson:72}
\BIBentryALTinterwordspacing
W.~H. Richardson, ``Bayesian-based iterative method of image
  restoration$\ast$,'' \emph{J. Opt. Soc. Am.}, vol.~62, no.~1, pp. 55--59, Jan
  1972. [Online]. Available:
  \url{http://www.osapublishing.org/abstract.cfm?URI=josa-62-1-55}
\BIBentrySTDinterwordspacing

\bibitem{kodinariya2013review}
T.~M. Kodinariya and P.~R. Makwana, ``Review on determining number of cluster
  in k-means clustering,'' \emph{International Journal}, vol.~1, no.~6, pp.
  90--95, 2013.

\bibitem{Huang16}
\BIBentryALTinterwordspacing
L.-C. Huang, M.~A. Neifeld, and A.~Ashok, ``Face recognition with non-greedy
  information-optimal adaptive compressive imaging,'' \emph{Appl. Opt.},
  vol.~55, no.~34, pp. 9744--9755, Dec 2016. [Online]. Available:
  \url{http://opg.optica.org/ao/abstract.cfm?URI=ao-55-34-9744}
\BIBentrySTDinterwordspacing

\end{thebibliography}

\newpage

\end{document}